\documentclass{article}

\usepackage[final]{neurips_2022_ml4ps}

\usepackage[utf8]{inputenc} 
\usepackage[T1]{fontenc}    

\usepackage{url}            
\usepackage{booktabs}       
\usepackage{amsfonts}       
\usepackage{hyperref}       
\hypersetup{
    colorlinks=true,
    linkcolor=blue,
    filecolor=magenta,      
    urlcolor=cyan,
    citecolor=blue,
}
\usepackage{url}  
\usepackage{booktabs}       
\usepackage{csquotes}
\usepackage{breakcites}
\usepackage{aligned-overset}
\usepackage{amsthm}
\usepackage{wrapfig}
\usepackage{diagbox}
\usepackage[toc,page,header]{appendix}
\usepackage{minitoc}
\usepackage{xspace}

\numberwithin{theorem}{section}

\newcommand{\pointnet}{PointNeXt\xspace}
\usepackage[most]{tcolorbox}
\tcbset{
    boxsep=0mm,
    boxrule=0pt,
    colframe=white,
    arc=0mm,
    left=0.5mm,
    right=0.5mm
}

\usepackage{xcolor}

\usepackage{thm-restate}
\newtheoremstyle{theoremdd}
  {\topsep}
  {\topsep}
  {\itshape}
  {0pt}
  {\bfseries}
  {. }
  { }
  {\thmname{#1}\thmnumber{ #2}\textnormal{\thmnote{ (#3)}}}
\theoremstyle{theoremdd}

\usepackage{nicematrix}

\usepackage{bm}
\usepackage{natbib}

\usepackage{outlines}
\usepackage{siunitx}
\usepackage{graphicx}
\usepackage{multirow}
\usepackage{caption}
\usepackage{subcaption}

\usepackage{cleveref}

\title{Cosmology from Galaxy Redshift Surveys\\with PointNet}

\author{
Sotirios Anagnostidis$^{*1a}$ \quad Arne Thomsen$^{*2b}$ \quad Tomasz Kacprzak$^{*2,3}$ \quad Tilman Tröster$^{*2}$\\[8pt]
\textbf{Luca Biggio$^{1,4}$ \quad Alexandre Refregier$^{2}$ \quad Thomas Hofmann$^{1}$}\\[16pt]
$^1$Department of Computer Science, ETH Zürich, CH \quad $^2$Department of Physics, ETH Zürich, CH\\
$^3$Swiss Data Science Center, Paul Scherrer Institute, CH \quad $^4$CSEM SA, Alpnach, CH\\
\texttt{$^a$sotirios.anagnostidis@inf.ethz.ch \quad $^b$athomsen@phys.ethz.ch}
}

\date{}

\begin{document}

\maketitle

\begin{abstract}
    In recent years, deep learning approaches have achieved state-of-the-art results in the analysis of point cloud data. In cosmology, galaxy redshift surveys resemble such a permutation invariant collection of positions in space.
    These surveys have so far mostly been analysed with two-point statistics, such as power spectra and correlation functions.
    The usage of these summary statistics is best justified on large scales, where the density field is linear and Gaussian.
    However, in light of the increased precision expected from upcoming surveys, the analysis of -- intrinsically non-Gaussian -- small angular separations represents an appealing avenue to better constrain cosmological parameters.
    In this work, we aim to improve upon two-point statistics by employing a \textit{PointNet}-like neural network to regress the values of the cosmological parameters directly from point cloud data.
    Our implementation of PointNets can analyse inputs of $\mathcal{O}(10^4) - \mathcal{O}(10^5)$ galaxies at a time, which improves upon earlier work for this application by roughly two orders of magnitude.
    Additionally, we demonstrate the ability to analyse galaxy redshift survey data on the lightcone, as opposed to previously static simulation boxes at a given fixed redshift.
\end{abstract}

\section{Introduction}
The standard model of cosmology contains a set of free parameters that relate to fundamental properties of the universe, such as the fraction of dark energy and dark matter, the dark energy equation of state, and the rate of cosmic expansion.
The matter density distribution evolves under the laws of gravity and cosmic expansion from a smooth, Gaussian field in the early universe, to a complex network of halos, filaments, and voids in the late universe.
The spatial distribution of galaxies traces the underlying dark matter density field. 
The cosmological parameters can be inferred by comparing the structures in this distribution from the observed survey data and those predicted by theoretical models \cite[see][and references therein]{Sanchez2017boss}.
Current cosmological analyses typically compress the information contained in the galaxy positions into two-point statistics, such as power spectra and correlation functions \citep{eboss2021completed}.
The two-point statistics are only sufficient for Gaussian random fields and therefore cannot fully capture the richer information contained in the galaxy positions. 

Using more powerful features, such as higher-order correlation functions~\citep{samushia2021information}, or other approaches~\citep{banerjee2021nearest,eickenberg2022wavelet}, either singularly or combined with each other, can improve performance.
Yet, the selection of an appropriate set of summary statistics is problem-dependent and prone to human bias.
This aspect motivates the adoption of tools to automatically extract and combine summary statistics in a meaningful way, i.e.\ to improve the prediction of cosmological parameters.
Aside from galaxy positions, other features can be added to increase the constraining power.
An example of such a property is the local density of the galaxy.
Marked correlation functions \citep{Satpathy2019marked,Philcox2021marked,Massara2022marked} report increased precision of measurements for various strategies to up-weight certain density regions.
This requires careful design of the marking strategy. 

Motivated by advancements in machine learning, recent work attempts to estimate parameters from the galaxy positions directly.
As an early example, ~\cite{ntampaka2020hybrid} frame the task as a 3D voxel regression, by applying an arbitrary scale partition of the input space into equally sized blocks.
A disadvantage of this approach is that the dynamical range of the density of galaxies makes the voxelisation either very lossy or computationally inefficient.
Furthermore, the permutation symmetries of the task are not respected.

A more natural representation is treating the data as a graph and employing graph neural networks (GNNs)~\cite{villanueva2022learning, makinen2022cosmic}, which are able to handle irregular and sparse data, via a message passing protocol defined between connected nodes.
However, this setting comes with two fundamental limitations.
Firstly, treating the data as a graph requires defining an adjacency matrix between the galaxies a priori.
Neighbors in this case are usually specified as points within a hard-defined threshold distance.
Secondly, these methods can scale poorly when increasing the available number of points, which is of major concern due to the relatively large galaxy catalogues which in the ongoing galaxy redshift surveys like DESI~\citep{DESI} can reach the order of $\mathcal{O}(10^6)$ objects.

We instead propose to treat the positional data directly in 3D space without making any further assumptions.
As a result, the underlying relationship between the positions can be learned automatically and no assumptions on an initial adjacency matrix are needed.
Point clouds are a natural representation of these geometric point sets, where an invariance to the order of the points within the set is desired.
Distances between points define a local neighborhood, whose properties should give good indications of the underlying data distribution.
We base our work on  PointNet~\citep{qi2016pointnet} and subsequent advances that learn features from point sets in a metric space.

An additional advantage of the analysis of PointNet is the fact that local density information can be naturally included in the input vector, extending the set of 3D positions $(x, y, z)$ with an additional local density proxy parameter, for example the halo virial mass $M$. 
Marked correlation functions use the local density values to optimally weight the correlation, and this weighting strategy needs to be carefully tuned. 
In contrast, the PointNet machine learning algorithm automatically finds the optimal weighting to maximize the information gain.

In this work, we demonstrate the potential of using PointNets for analysing redshift surveys with a large number of galaxies, reaching $\mathcal{O}(10^4) - \mathcal{O}(10^5)$ objects, with prospects of further increase when ran on larger hardware.
We show that the local density information can be straightforwardly included in the data vector, extending the set of galaxy properties by one additional parameter.
We compare the accuracy of PointNets and two-point statistics in predicting the underlying cosmological parameters and show that our approach performs favourably, especially when the additional feature $M$ is included.

\section{Methodology}
\begin{figure}[!t]
    \centering
    \includegraphics[scale=0.55]{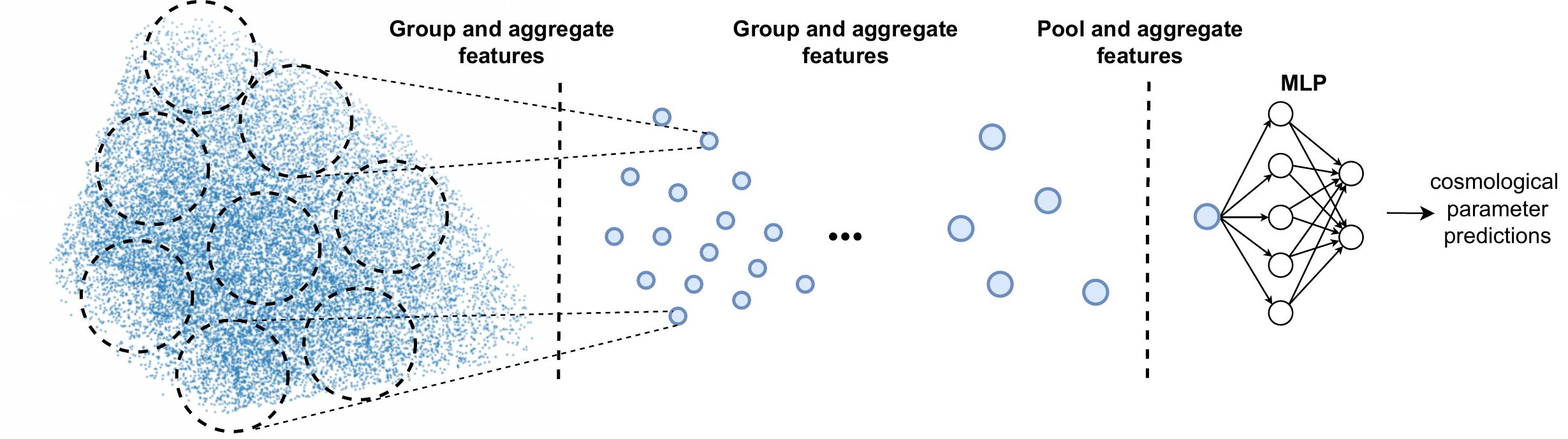}
    \caption{\textbf{Schematic illustration of the proposed approach.} For galaxy redshift surveys, cosmological parameter inference is traditionally performed by manually extracting summary statistics, i.e. features, from a set of points in 3D coordinates. Our approach differs in that relevant features are automatically extracted by a hierarchical PointNet directly processing the point cloud and outputting the corresponding cosmological parameters. This way, the need for introducing hand-crafted summary statistics can be entirely avoided.}
    \label{fig:pointnext}
\end{figure}

Our dataset is derived from the dark matter halo catalogs contained in the \textsc{CosmoGridV1}~\citep{cosmogrid, Fluri2022}, which contains dark-matter-only N-body simulations varying the six cosmological parameters $\Omega_m$, $\sigma_8$, $w_0$, $H_0$, $n_s$ and $\Omega_b$ of the $w$CDM model.
The distribution of halos is primarily affected by the parameters $\Omega_m$, the present-day fraction of matter in the Universe, and $\sigma_8$, a measure of the variance, or clumpiness, of the matter distribution.
We, therefore, restrict the present study to these two parameters.
The remaining ones are implicitly marginalized over in the analysis.
To simplify the physics modeling, we use dark matter halo positions as a proxy for galaxy positions and work in physical space, rather than redshift space.

The positions of the halos define a permutation-invariant set of points in 3D comoving coordinates.
We assume that the distance metric between these points defines a local neighborhood that encodes properties of the underlying data generation process.
This concept of local neighborhood allows for the capture of attributes that may not be uniformly distributed across the set of available points, which is in contrast to the global view two-point functions provide.
Therefore, our basic intuition is to exploit the local structure and learn fine-tuned spatial features based on these neighborhood attributes.
In addition, we include the halo mass $M$ as a proxy for the local density in selected experiments. Then, the input to the neural net is a set of positions and mass: $(x, y, z, M)$.

Based on recent work on hierarchical neural network point set learning~\citep{qi2017pointnet++, zhao2021point}, we partition the set of halos on overlapping local regions by the distance metric of the underlying space, which is Euclidean.
By aggregating predictions across these regions on increasingly large scales, we progressively expand the receptive field and produce higher-level features, see also Fig.~\ref{fig:pointnext}.
This enables us to efficiently train on large sets of point clouds, which capture both local and global information.

\section{Experiments}
\label{sec:experiments}

\begin{figure}[!tb]
    \begin{subfigure}{0.59\linewidth}
    	\centering
        \includegraphics[width=0.9\linewidth]{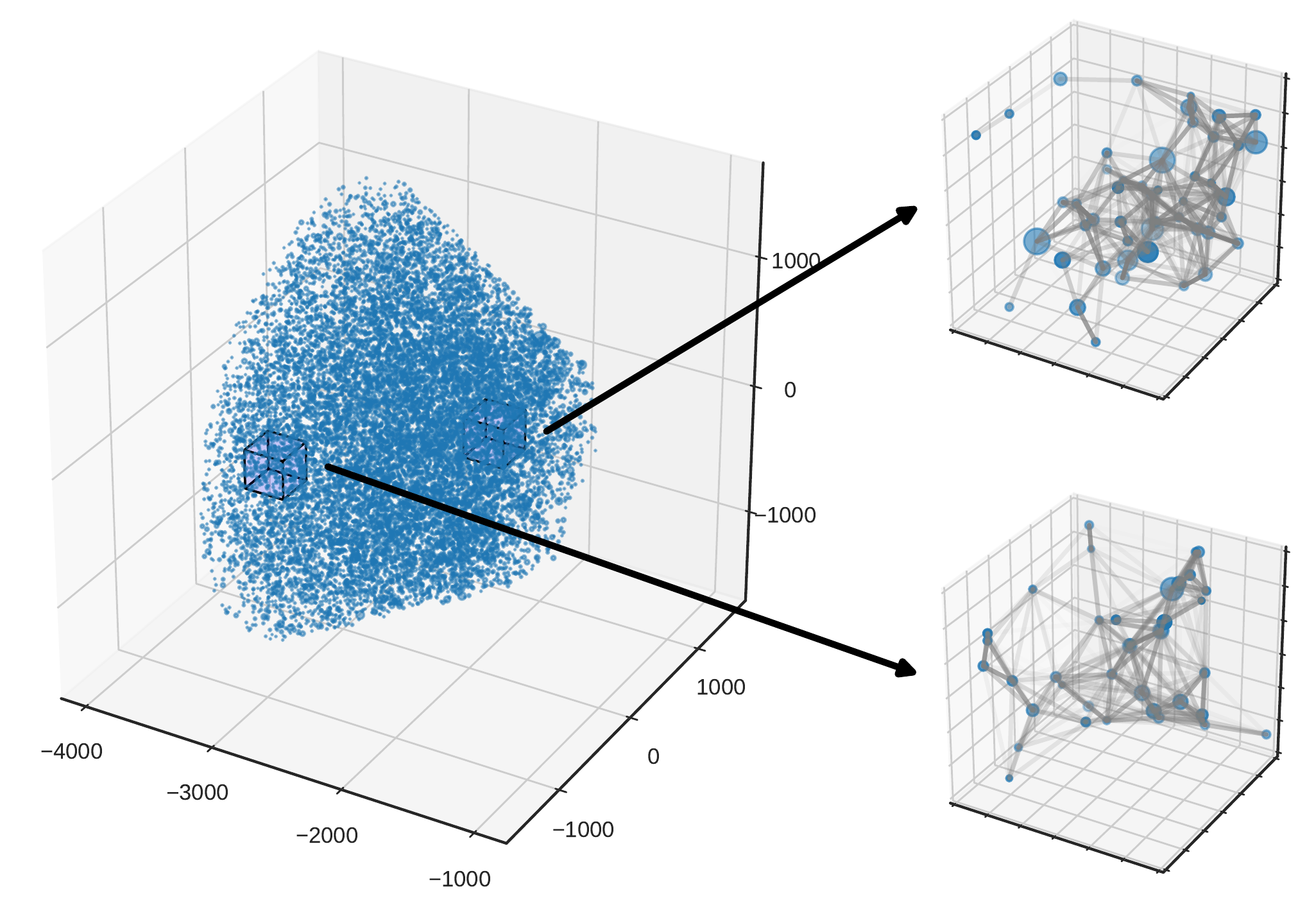}
        \caption{\textbf{Example point cloud of dark matter halos} on a lightcone, where the size of the points is proportional to their virialized mass $M$. The axes give the comoving coordinates in units of Mpc and $1.6\times 10^4$ points are depicted in total. The two subfigures include boxes of size 300 Mpc. There, the edge-strength corresponds to the inverse of the distance between points.}
        \label{fig:pointcloud}
	\end{subfigure}
	\begin{subfigure}{0.4\linewidth}
		\centering
		\includegraphics[width=0.9\linewidth]{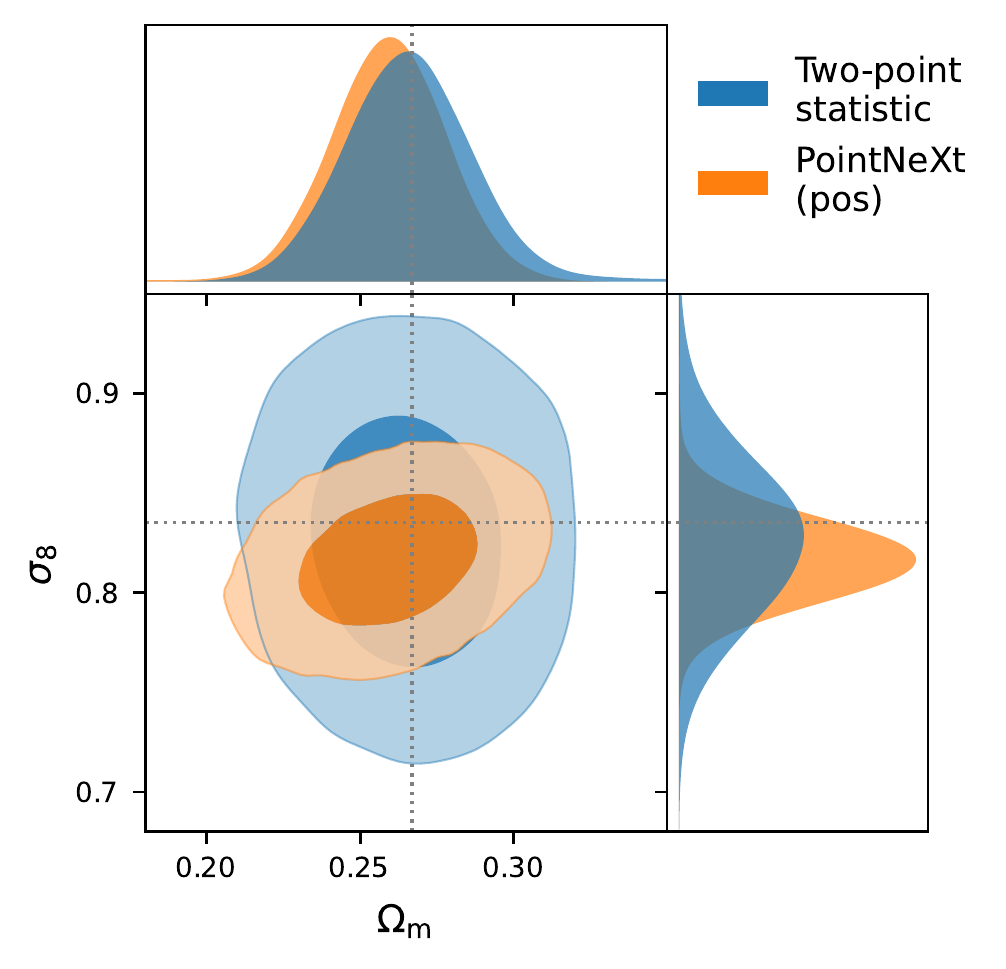}
		\captionsetup{width=0.9\linewidth}
		\caption{\textbf{Comparison of the cosmological constraints.} The contours correspond to the posterior distribution of $\Omega_m$, the fraction of matter in the Universe, and $\sigma_8$, the variance of the matter field. \pointnet outperforms the two-point statistic for equal numbers of points, which is set to $3.2 \times 10^4$.}
		\label{fig:corner_plot}
	\end{subfigure}
\end{figure}

\textbf{Data.} The survey volume encompasses a round patch of 1000 square degrees, which is roughly $2.5\%$ of the full sky and a redshift range of $z \in [0.4, 1]$, see Fig.~\ref{fig:pointcloud} for a visualisation.
The dark matter halos create a lightcone, where the properties of the density field evolve in redshift; in effect, the higher $z$ halos are clustered the way they were at the time in cosmic history corresponding to that redshift.
The number of dark matter halos contained in this volume is cosmology-dependent.
For consistency, we only consider point clouds of a fixed number of points (see Table~\ref{tab:incremental_results}) and discard cosmologies with fewer points, which are only found for the lowest values of the $\sigma_8$ parameter.
These are orders of magnitude more points than in previous analyses employing GNNs, which was $\mathcal{O}(10^2)$ in~\cite{makinen2022cosmic}, and $\mathcal{O}(10^3)$ in~\cite{villanueva2022learning}.

The \textsc{CosmoGridV1} is made up of 2500 unique combinations of the six cosmological parameters distributed according to a Sobol sequence.
The suite contains seven independent N-body simulation runs for each such grid point, which result from differing random initial conditions and each cover the full sky.
We cut out 12 non-overlapping survey patches from each of these to arrive at 84 independent realizations of the point cloud per cosmology.
Therefore, the dataset encompasses a total of 210\,000 elements before the removal of cosmologies which only contain an insufficient number of halos.
We split the original data into training ($90\%$) and testing ($10\%$).

We obtain consistent comoving coordinates throughout the grid of cosmologies by fixing a fiducial cosmology with respect to which the observable angular and redshift coordinates are converted.
In addition to the comoving coordinates $(x,y,z)$, the catalogs contain the halo mass $M$.

\textbf{Point cloud networks.} We build upon the recent innovations in point cloud processing by neural networks in the PointNet family~\citep{qi2016pointnet}.
These are based on the idea that the density of point sets is not uniform across the receptive field, hindering performance when methods disregard this fact.
PointNet++~\citep{qi2017pointnet++} and succeeding improvements proposed in PointNeXt~\citep{Qian2022}, learn layers in an adaptive fashion, from multiple scales.
We adopt the latter of these methods, which was recently shown to achieve state-of-the-art results across a variety of point cloud applications.

\textbf{Two-point statistics.} We use isotropic two-point correlation functions as the baseline for our experiments and estimate them using the \textsc{TreeCorr}~\cite{treecorr} library. 
The correlation functions are calculated between the same numbers of points as the ones that make up the point clouds.
We compress these two-point feature vectors using a simple multilayer perceptron (MLP) regressor, allowing for a more direct comparison to the \pointnet models, as well as enabling efficient conditional density estimation. 
For further details, see Appendices~\ref{sec:two-point-appendix} and \ref{sec:inference-appendix}.

\begin{table}[!b]
    \centering
        \begin{tabular}[b]{c|ccc}
            \toprule
            \# points & \pointnet (pos) & \pointnet (pos+$M$) & Two-point statistic \\
            \midrule
            $0.8 \times 10^4$ & $3.6 \times 10^{-3}$ & $1.3 \times 10^{-3}$ & $8.3 \times 10^{-3}$\\
            $1.6 \times 10^4$ & $2.4 \times 10^{-3}$ & $6.7 \times 10^{-4}$ & $3.4 \times 10^{-3}$\\
            $3.2 \times 10^4$ & $1.3 \times 10^{-3}$ & $5.8 \times 10^{-4}$ & $1.8 \times 10^{-3}$\\
            \bottomrule
        \end{tabular}
        \vspace{1em}
        \caption{\textbf{Comparison of the MSE for our \pointnet and an MLP trained on two-point correlations.} The MSE is calculated on a test set of predictions for both $\Omega_m$ and $\sigma_8$. The results reported in the third column are obtained with the halo mass as a feature additional to their position.}
        \label{tab:incremental_results}
\end{table}
\textbf{Results.} We consider two metrics to assess the performance of our methods: the mean-squared error (MSE) on the test set and the posterior distribution of the cosmological parameters with respect to a fixed mock observation.

The MSE of \pointnet and the two-point statistic is shown in Table~\ref{tab:incremental_results}.
The results indicate that \pointnet obtains more precise predictions for the same number of halos. 
The difference between the two methods is larger for smaller numbers of points, which implies that the PointNets are more robust to noise.
More striking is the boost in performance the inclusion of the mass $M$ as an additional feature brings. 
For the PointNets, the MSE drops by more than a factor of two with this additional information. 

To estimate the posterior distributions of cosmological parameters, we perform conditional density estimation on the regressed parameters from \pointnet and the compressor of the two-point statistics.
We refer the reader to appendix \ref{sec:inference-appendix} for details on the posterior inference.
In Fig.~\ref{fig:corner_plot}, we visualize the resulting parameter posteriors when using $3.2\times10^4$ points.
In line with what is observed for the MSE, \pointnet is able to constrain the parameters more precisely than the two-point statistic, especially the parameter $\sigma_8$, which describes the variance of the matter field.
While the MSE of the \pointnet model is markedly improved by the inclusion of the mass $M$ as a feature, we find that this is mainly due to a reduction of outliers.
The size of the $(\Omega_m, \sigma_8)$ posterior is however not strongly affected by the inclusion of the halo mass as a feature.

\section{Discussion}

For our experimental setting, the standard two-point correlation function does not offer any further potential for improvement beyond the results reported in \Cref{tab:incremental_results}, thus reaching its limits.
While marked functions could be used to incorporate the evidently beneficial mass information into the correlators, these require careful modeling choices.
In contrast, the PointNets automatically learn how to optimally handle the information contained in this additional feature, which makes them especially suited for the task.
As in this study we have not performed an extensive optimization of the PointNets both in terms of hyperparameter tuning and the selection of architecture, we anticipate that the gap in performance between the methods in \Cref{tab:incremental_results} can be increased, even when considering only the galaxy positions.
Similarly, the size and complexity of the PointNet can be increased on larger hardware, thus potentially leading to more precise parameter measurements.
Another avenue of improvement is to utilize models that ensure the adherence to physical symmetries like rotation or translation equivariance, as this was shown to improve the results in~\cite{makinen2022cosmic}.

\section{Conclusion and Outlook}

In this work, we use PointNets to directly regress cosmological parameters from simplified simulated galaxy redshift surveys.
We show that for an equal number of points, our PointNet implementation outperforms the de facto standard approach, the two-point correlation function.
We show that the improvement is twofold:
Firstly, the PointNets obtain better results when acting on the positions only, presumably by extracting non-Gaussian information the correlators aren't sensitive to.
Secondly, the inclusion of the mass as an additional feature improves their performance in terms of the MSE, while only marignally improving the constraints.
Given the ease of including more features to the point vector, this is a promising result. 
We leave the investigation of how it can be used to gain constraining power to future work.

With this in mind, it is a notable advantage of the PointNets that additional features can be incorporated directly and without the need for fine-tuning, which is not the case for extensions to the correlation function.
Therefore, a promising direction for follow up work is to apply the method to point clouds made up of galaxies instead of dark matter halos, as more physical features apart from the positions and mass can be modeled.
For example, the inclusion of measurements of galaxy magnitudes and colors can help constrain galaxy evolution models in connection to their density environment.

Our implementation of PointNets can currently handle $\mathcal{O}(10^4) - \mathcal{O}(10^5)$ galaxies at a time, which marks an improvement of roughly two orders of magnitude compared to earlier work deploying deep learning methods to this application.
As there is still an order of magnitude gap to the $\mathcal{O}(10^6)$ galaxy positions mapped in ongoing state-of-the-art galaxy redshift surveys, scaling our method to even larger numbers of points is another direction for future extensions of this work.
We expect that this can be achieved, for example, by using the latest large memory, multi-GPU nodes.

\section*{Impact statement}\label{impactstatement}
Upcoming galaxy surveys are expected to provide a wealth of new data that can shed light on the nature of dark matter and dark energy. 
The design of new tools and techniques to process such data is of paramount importance to this task.
The presented approach contributes to this by demonstrating a simulation-based inference analysis framework with PointNets. They can efficiently operate on large galaxy point clouds, naturally incorporate more galaxy properties, and maximize the information gain form the data.
We do not foresee any ethical concerns with this work.

\bibliographystyle{abbrvnat}
\bibliography{main}

\newpage
\appendix

\section{Experimental Setup}
\label{sec:implementation_details}

\subsection{Point cloud networks}
More details regarding our implementation setup are displayed in Table~\ref{tab:hyperparameters}. We highlight that we mainly follow the experimental settings of~\cite{Qian2022}. For our most computational intensive experiments we train on 8 NVIDIA Tesla P100 GPUs.

\begin{table}[h]
\centering
\begin{tabular}{cl|c}
  \hline
  \rule{0pt}{1.\normalbaselineskip} & \multicolumn{1}{c|}{\textbf{Hyperparameters}} & \multicolumn{1}{c}{\textbf{Value}}\\[1mm]
  \hline
  \rule{0pt}{1.\normalbaselineskip}\parbox[t]{2mm}{\multirow{6}{*}{\rotatebox[origin=c]{90}{{Architecture}}}} & radius & 0.1 \\
  & strides & [4, 4, 4, 4] \\
  & blocks & [4, 4, 7, 4] \\
  & width & 32 \\
  & expansion & 4 \\
  & nsample & 32 \\[1mm]
  \hline
  \rule{0pt}{1.\normalbaselineskip}\parbox[t]{2mm}{\multirow{6}{*}{\rotatebox[origin=c]{90}{Training}}} & Learning rate & $3\epsilon^{-4}$ \\
  & Learning rate scheduler & `cosine' \\
  & Adam $(\beta_1, \beta_2)$ & (0.9, 0.999) \\
  & Batch size & 64 \\
  & Weight decay & 0.0001 \\
  & Loss & `mse-loss' \\[1mm]
  \hline
  \rule{0pt}{1.\normalbaselineskip}
\end{tabular}
\caption{Hyperparameters for the PointNeXt experiments.}
\label{tab:hyperparameters}
\end{table}

\subsection{Two-point statistics}
\label{sec:two-point-appendix}
We fix the number of randoms in the Landy-Szalay estimator~\cite{landy_szalay} to $2.56 \times 10^5$ in all cases.
The summary statistic leaves freedom to choose appropriate binning, which controls both the scales included (via the bin limits), and the amount of noise (via the bin widths). 
We fix these parameters such that we obtain 50 linearly spaced bins in the range from 0 Mpc to 300 Mpc.

\subsection{Cosmology parameter inference}
\label{sec:inference-appendix}
We wish to derive posterior distributions on the cosmological parameters, similar to what is being done in traditional cosmological analyses.
Since the likelihood of the halo distribution is intractable, we make use of simulation-based inference \cite{Alsing2019,Cranmer2020}.
The high dimensionality of the input data, the halo catalog, makes the direct application of conditional density estimation difficult.
We therefore compress the input data to a low-dimensional representation on which we can then perform conditional neural density estimation.
Different choices as to how one can compress the high-dimensional data have been proposed for cosmological data sets \cite{Charnock2018, Fluri2019, Jeffrey2021, Fluri2021}.
Here we consider three approaches: using the cosmological parameters as summaries, together with a simple MSE loss; using the cosmological parameters as summaries while also predicting their covariance, together with a likelihood loss, similar to \cite{Fluri2019}; and using summaries that maximise the mutual information with the true cosmological parameters, similar to \cite{Jeffrey2021}. 

We assess the efficiency of these three compression schemes on the two-point statistics due to the lower computational cost compared to the \pointnet model. 
We use an MLP with two hidden layers of width $n_\mathrm{hidden}=128$ as the compressor, trained for 200 epochs, with the learning rate decreasing from $4\times 10^{-4}$ every 40 epoch by a factor of two. 
We find that the differences between these three approaches when compressing the two-point statistics are negligible, both in terms of the MSE, as well as the inferred posterior distributions. 
We therefore use the cosmological parameters as summaries, together with a MSE loss, as our fiducial choice for both the two-point data and \pointnet.

\begin{figure}
    \centering
    \includegraphics[width=0.6\linewidth]{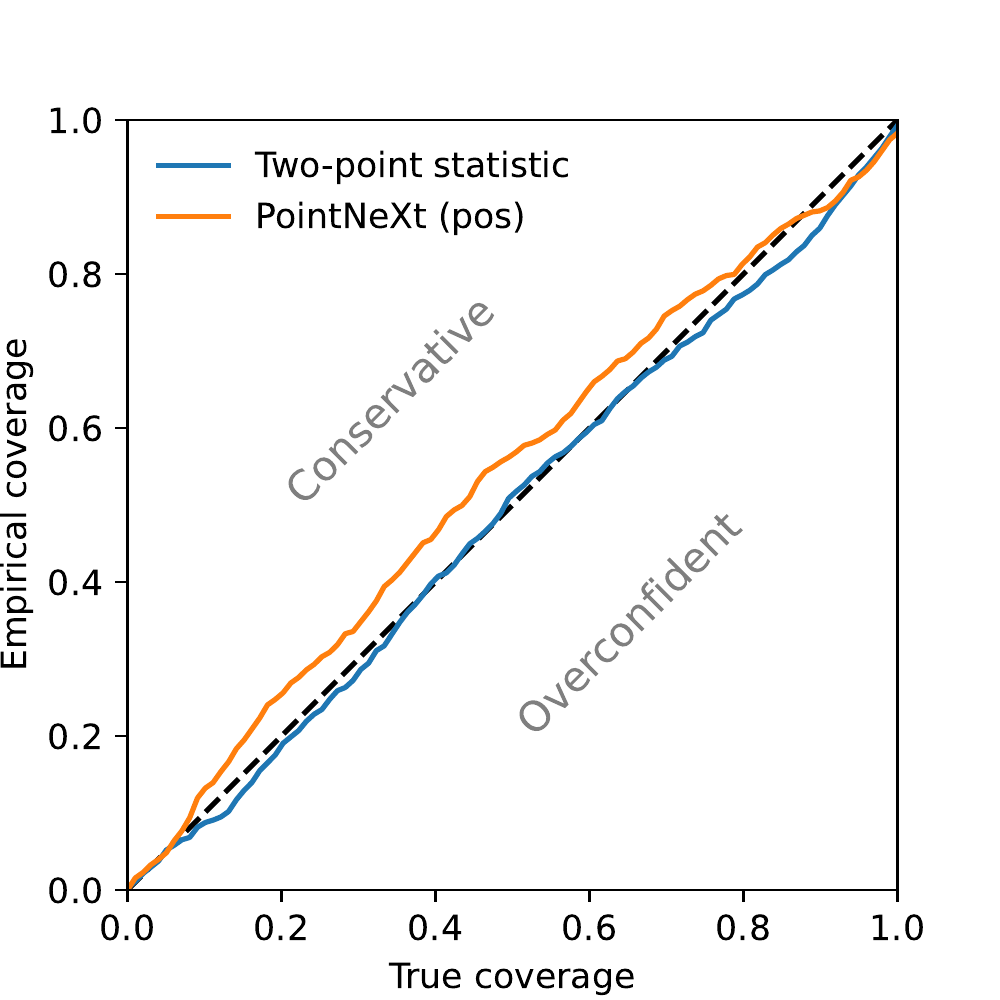}
    \caption{\textbf{Expected coverage of the posteriors in Fig.~\ref{fig:corner_plot}.} A perfectly calibrated posterior has an empirical expected coverage probability equal to the nominal expected coverage probability, corresponding to a diagonal line. 
    Conservative posteriors lie above the diagonal, while overconfident posteriors lie below it.}
    \label{fig:eecp_plot}
\end{figure}

We use a three-component Gaussian mixture model for the conditional density estimation of the summaries. 
The posterior distributions are then obtained by conditioning the density estimator on a single representative sample of the summaries.
The coverage of the estimated posterior distributions are validated using the empirical expected coverage probability \cite{Hermans2021} and are found to be well calibrated, as shown in Fig.~\ref{fig:eecp_plot}.

As a sanity check we also approximate a Gaussian likelihood for the two-point statistics by considering the mean over all realisations at each grid point as the model and using the realisations at the fiducial cosmology for the data covariance matrix.
This gives a noisy and sparse estimate of the likelihood over the whole prior volume. 
The resulting posterior distribution agrees with that of the compressed two-point statistics but we refrain from a quantitative comparison due to the sparsity of the posterior samples and noise of this Gaussian likelihood approximation.

\end{document}